\documentclass[letterpaper,12pt]{article}
\usepackage[hmargin=1in,vmargin=1in]{geometry}
\usepackage{natbib}
\usepackage{setspace,amssymb,amsmath,graphicx,wasysym,fancyhdr}

\begin{document}
\title{The duration of load effect in lumber as stochastic degradation}
\author{Samuel W K Wong$^1$ and James V Zidek$^2$\\
$^1$  Department of Statistics, University of Florida, Gainesville, FL\\
$^2$  Department of Statistics, University of British Columbia, Vancouver, BC\
}
\maketitle{}

\begin{abstract}
This paper proposes a gamma process for modelling the damage that accumulates over time in the lumber used in structural engineering applications when stress is applied. The model separates the stochastic processes representing features internal to the piece of lumber on the one hand, from those representing  external forces due to applied dead and live loads.  The model applies those external forces through a time-varying population level function designed for time-varying loads. The application of this type of model, which is standard in reliability analysis, is novel in this context, which has been dominated by accumulated damage models (ADMs) over more than half a century. The proposed model is compared with one of the traditional ADMs. Our statistical results based on a Bayesian analysis of experimental data highlight the limitations of using accelerated testing data to assess long-term reliability, as seen in the wide posterior intervals.  This suggests the need for more comprehensive testing in future applications, or to encode appropriate expert knowledge in the priors used for Bayesian analysis.

\end{abstract}

\newpage

\section{Introduction}\label{sect:introduction}
Wood placed under a sustained load over a period of time in an engineering application will sustain damage due to the ``duration of load'' (DOL)  effect \citep{karacabeyli1991state}.  The importance of this effect led to the development of models for predicting it so that it could be incorporated into the establishment of design values and safety factors in applications. However shortcomings in these models that we now describe led to the development by the authors of the alternative  approach described in this paper.  The two approaches are then compared in an illustrative application.

Long term damage in a piece of lumber, resulting in a reduction of its strength-bearing capacity, depends on the load level and how it is applied (e.g., via bending, compression, tension). The speed at which a piece weakens over time may also depend on a combination of factors such as the visco--elasticity of the wood, temperature and moisture.  Even a relatively small constant load applied over a sufficiently long period of time may lead to failure (known as creep rupture).  According to the review of \citet{rosowsky2002another}, the effect was first recognized by \citet{haupt1867general}. But it does not seem to have been formally incorporated into design standards until \citet{wood1960relation} produced the so--called Madison Curve for doing so. The curve is still in use today for estimating the DOL effect on the strength of wood.

However the purely empirical approach of Wood led only to a fitted curve. So an alternative dynamic model was developed to describe how damage accumulated over time as a function of the stress load profile \citep{barrett1978durationI,barrett1978durationII,gerhards1979time}. These accumulated damage models (ADMs) differ in detail, but the idea is the same, to focus on the rate at which damage accumulates rather than the damage itself, using an ordinary differential equation (ODE)
\begin{equation}\label{eq:dynamicmodel}
{ d \alpha (t)\over dt } = F[\alpha(t), \sigma(t), \mathbf{\phi}],
\end{equation}
which represents the rate of damage accumulation for a randomly selected piece of lumber.  The vector $\mathbf{\phi}$ contains (random) parameters associated with the piece itself, with their joint probability distribution depending on population parameters that must also be fitted to implement the model.  Once a piece is selected, the model (\ref{eq:dynamicmodel}) deterministically describes the rate at which damage accumulates in that piece.

While the accumulated damage at time $t$, $\alpha(t)$, is unobservable, the ODE provides a framework onto which the other elements of the model can be attached. It is calibrated so that $\alpha(t)=0$ when $t=0$ and no damage has occurred, and $\alpha(T_l)=1$ at time $t=T_l$ when the piece fails. Here  $\sigma(t) = \tau(t)/\tau_s$,  where $\tau(t)$ (psi) is the applied stress at time $t$ and $\tau_s$ (psi) is the `short-term breaking strength' of the piece (commonly defined to be the stress at which the piece would fail were it to be subjected to a ramp load test of duration $\sim$1 minute).

For definiteness, this paper will focus on a representative and well--known ADM, the ``Canadian model''  proposed by \cite{foschi1984reliability}. That model is based on the two-term approximation obtained from a Taylor expansion of $F$ in Equation (\ref{eq:dynamicmodel}) as a function of $\alpha(t)$, namely
\begin{equation}\label{eq:canadianmodel}
  \dot{\alpha}(t) = a [\tau(t) - \sigma_0 \tau_s]_+^b
  + c [\tau(t) - \sigma_0 \tau_s]_+^n \alpha(t)
  \end{equation}
where $a$, $b$, $c$, $n$, $\sigma_0$ are log--normally distributed random effects for the piece.  Here $\sigma_0$ is known as the stress ratio threshold, and $x_+ = \max(x,0)$.  Thus in this model, no damage accumulates in the piece when $\tau(t) < \sigma_0 \tau_s$.

However \citet{ellingwood1991duration} point out that the Canadian model cannot be nondimensionalized.  That is a serious issue, since a model that represents a natural process cannot ultimately depend on how the quantities involved in the model are measured.  The concern was seen to be of sufficient importance that \citet{ellingwood1991duration} exclude the model from their comparative analysis of ADMs. The review of ADMs in \citet{hoffmeyer2007duration} instead modifies the Canadian model to correct the dimensions; this difficulty of the Canadian model, as well as in other ADMs, was also described in \citet{zhai2011dynamic} and \citet{zhai2012review}.  \citet{wong2016dimension} also address the problem by invoking the Buckingham $\pi$ theorem to build reparametrized models that are dimensionally consistent while retaining their functional form.

Another difficulty associated with the ADM approach is  its computational burden due to the need to solve ODEs such as Equation (\ref{eq:canadianmodel}) numerically for each individual piece of lumber, one that restricts the use of standard likelihood-based methods for their analysis.  As a result, uncertainties in both the parameter estimates and subsequent reliability calculations are difficult to quantify.  \citet{yang2017adm} address this latter difficulty by proposing approximate Bayesian computation techniques to perform the analysis on a solid statistical platform; however, a large cluster of CPUs is needed to carry out the subsequent reliability calculations with high--accuracy ODE numerical solvers.

As a final limitation of the ADM approach, randomness in the process of damage accumulation within a given piece is ignored, which may not be realistic.  In consequence, estimates of ADM parameters are difficult to interpret as population level and piece--specific modeling are inherently intertwined.  So this  paper presents a new approach for modelling the DOL effect that overcomes the difficulties described above. It is based on the gamma process, a standard approach to modelling degradation that has a long history \citep{lawless2004covariates}. But it has not previously been used for wood products as far as the authors are aware.  To successfully apply this approach to lumber, we formulate a model for the time-varying shape parameter that accounts for the time-varying loads which pieces must sustain.

In a major point of departure from the ADM approach above, the degradation of a piece of lumber under the gamma process  remains random (even conditional on it having being selected):  it is represented by a stochastic process $Y_t,~t\geq 0$, that describes the damage accumulated up to time $t$. That process is internal to the piece, and can be thought of as representing its random progress of damage. The future combination of dead and live loads, which may be a random process, are external to that piece.  Given a realized load profile, these two ingredients are fused through the deterministic time--varying population level shape parameter. This separation of internal and external sources of variability has advantages in terms of the interpretability of the results and facilitates the use of principled statistical methods of analysis.

The paper presents a number of notable findings that we now summarize.
\begin{itemize}
	\item Weak evidence is found of a threshold effect below which no degradation in the population occurs, that being an estimated threshold level of 413psi for the population from which the test data in our illustrative application are drawn. But the posterior credibility bands are wide, making the possibility of no threshold quite plausible.
	\item The experimental data suggest that the degradation as a function of time for a lumber population cannot be explained by the simple power law commonly used in gamma process applications.
	\item Our reliability analysis suggests that under a simulated future dynamic occupancy load, the chance of failure of a piece of lumber before the end of fifty years is 9\%. This contrasts with a more optimistic estimate 1.5\% obtained by application of an ADM under the same simulated future.
	\item Finally, the analysis of future residual life of a piece of lumber that has survived at least four years under a constant load of 3000psi has a long right-tailed distribution with a median survival time in the range of 22 to 333 years if that same load is sustained indefinitely.  That substantial amount of uncertainty revealed by our analysis, points to the need for much more testing to precisely estimate reliability under low sustained loads.
\end{itemize}

The paper is organized as follows. Section \ref{sect:specimen} introduces the gamma process as a way of describing the damage due to the stress applied to wood when placed in service. Section \ref{sect:degradation} describes how the gamma process may be used to model degradation. In particular   it is shown how the random degradation process for a randomly chosen piece of lumber gets coupled to a model for its population and how the applied stress profile operates to cause damage to accumulate. A major contribution follows next in Section \ref{sect:BayesianAnalysis}: a Bayesian approach for applying the gamma process model is developed and applied to data obtained in an accelerated testing experiment designed to explore the duration of load effect.   A discussion about the lessons learned in Section \ref{sect:BayesianAnalysis} about the reliability of lumber comes next in Section \ref{sect:reliabilityanalysis}.  Another application follows in Section \ref{sect:residuallife}; there it is shown how the residual life of a piece of lumber in service can be predicted. Further discussion and concluding remarks follow in Section \ref{sect:discussion}.

\section{The gamma process as a specimen--specific stochastic model}\label{sect:specimen}

In this section we briefly review the basics of the gamma process as it relates to modeling lumber degradation.

Let $Y_t\geq 0$ be the stochastic process representing the accumulated damage (or degradation in the terminology of reliability theory) in a piece of lumber at time $t$.  Assume $Y_0=0$ and that $Y_t\geq 0$ is nondecreasing over time, as any damage sustained is irreversible.  We say that the piece reaches a state of failure at time $t=T$ when the damage exceeds a pre-specified threshold level indicating failure.  Without loss of generality, we may scale the degradation process so that failure occurs at $Y_T=1$.  Virtually, the degradation process can be thought to continue for $t>T$ even though by that time the specimen will have failed.

Conditional on the parameters for a randomly selected lumber specimen, assume $Y_t,~t\geq 0$ has stochastically independent increments, i.e.~for any sequence of times $t_1< \dots < t_n$, the increments
$Y_{t_i} - Y_{t_{i-1}},~i=1,\dots n$ are stochastically independent. The distribution of these increments may depend on factors internal to the specimen as well as the external effects of the applied stress, resulting in damage that accumulates as a series of successive jumps of random size.  The particularly simple family of models we adopt assumes $Y_t$ is a compound Poisson process with intensity function $\lambda_t,~t>0$, i.e.
\[
Y_{t_i} - Y_{t_{i-1}} = \sum_{i=1}^{
N_{
(t_{i-1},t_{i})
}
} X_i
\]
where conditional on the model parameters
\begin{eqnarray*}
	P[N_{
(t_{i-1},t_{i})
} = n] &=& \frac{\Lambda_{
(t_{i-1},t_{i})
}^n \exp\{-{
\Lambda_{
(t_{i-1},t_{i})
}
}
\}
}{n!},~{\rm with}\\
	\Lambda_{
(t_{i-1},t_{i})
} &=& \int_{t_{i-1}}^{t_i} \lambda_t dt\\
\end{eqnarray*}
while the random jumps $X_i$, which are independent of the Poisson count process, have a gamma distribution with shape parameter $\eta$ and scale $\xi$.  Standard theory then implies that conditional on the model parameters
\begin{eqnarray*}
	E[Y_{t_i} - Y_{t_{i-1}}]&=&  \xi \eta\Lambda_{
(t_{i-1},t_{i})
},~{\rm while} \\
	Var[Y_{t_i} - Y_{t_{i-1}}]	&=&
 \xi \eta\Lambda_{
(t_{i-1},t_{i})
} [ \eta + \xi].	\\	
\end{eqnarray*}

As the intensity parameter increases and the gamma shape parameter decreases,  we approach in the limit, the so--called gamma process that has an infinite number of infinitesimally small jumps. That model has been used extensively to model degradation. More formally
\[
Y_{t_i} - Y_{t_{i-1}}\sim
Ga[\xi,\eta_{t_i}- \eta_{t_{i-1}}]
\]
where $\eta_t \ge 0$ is a nondecreasing function, and $Ga[\xi,\eta_t]$ denotes the gamma distribution with scale parameter $\xi$ and shape parameter $\eta_t$.
The scale $\xi=\xi(x)$ is a scalar-valued quantity that could also depend on fixed covariates $x$ associated with a specimen, such as the modulus of elasticity.  From standard theory we then obtain
\begin{eqnarray*}
E[Y_t \mid \xi,\eta_t] &=& \xi(x)\eta_t\\
Var[Y_t \mid \xi,\eta_t] &=&
\xi(x)^2\eta_t,{\rm and}\\
CV[Y_t \mid \xi,\eta_t] &=& \eta^{-1/2}.	
\end{eqnarray*}

Provided that multiple gamma processes have
the same scale $\xi$, which in effect means each has a scale that is a known multiple of $\xi$, their sum is also a gamma process. More precisely, assume that conditional on $\eta_{it}$ and $\xi$, the $\{Y_{it}\}$, $i=1,\dots r$
are independent gamma processes with shapes $
\eta_{it}$ and scales $\xi$. Then the sum
\begin{equation}\label{eq:multiple}
	Y_t = Y_{1t} +\dots +  \dots Y_{rt}
\end{equation}
is also a gamma process with shape $\eta_t = \sum_i \eta_{it}$ and scale $\xi$.

This is a useful property since it provides a convenient framework for combining different processes that contribute to degradation.  The process corresponding to damage due to the applied load profile is that of primary interest in this paper.  As potential extensions, other external factors that contribute to damage such as the time-varying moisture content and temperature of the environment could be incorporated as separate components in Equation (\ref{eq:multiple}).   However, we do not at present have the data to illustrate these refinements to the model.

\section{Degradation to failure}\label{sect:degradation}

\subsection{Probability distribution of failure time}

The gamma process induces a probability distribution of failure times $T$, which we briefly review as follows.  Detailed proofs of these results can be found in \citet{paroissin2014failure}.  The survival function for $T$ is
\begin{eqnarray}\label{eq:surv}
P[T > t \mid \xi,\eta_t] = P[Y_t \leq 1 \mid \xi,\eta_t] = \int_{0}^{1}\frac{u^{\eta_t -1}e^{-u/\xi}}{\xi^{\eta_t}\Gamma[\eta_t]} \, du =1 -  \frac{ \Gamma(\eta_t, 1/\xi)}{\Gamma(\eta_t)},
\end{eqnarray}
where $\Gamma(\cdot,\cdot)$ denotes the upper incomplete gamma function.  When $\eta_t$ is differentiable, it follows that the probability density of $T$ needed for the construction of the likelihood function is
\begin{eqnarray}\nonumber
f_T[t \mid \xi,\eta_t] &=&
- \frac{d P[T > t \mid \xi,\eta_t]}{dt}
\\&=& \label{eq:likT}
\dot{\eta}_t \left( \Psi(\eta_t) - \log (c/\xi) \right) \left( 1 -  \frac{ \Gamma(\eta_t, 1/\xi)}{\Gamma(\eta_t)} \right) \\\nonumber
&\mbox{}& + \frac{\dot{\eta}_t}{\eta_t^2 \Gamma(\eta_t)} (c/\xi)^{\eta_t}\, _{2}F_{2} (\eta_t, \eta_t ; \eta_t+1, \eta_t+1; -1/\xi),
\end{eqnarray}
where $\Psi$ is the digamma function and $_{p}F_{q}$ is the generalized hypergeometric function of order $p,q$.

\subsection{Damage due to load applied}

Of primary interest is characterizing the gamma process representing damage due to the stress applied.  Suppose the load profile over time with which the population is stressed, $\tau(t); t \ge 0$ is given.  Then $\tau(t)$ has a fundamental role in determining the corresponding value of $\eta_t$.  In particular, $\eta_t$ must account for the degradation effects of the entire load history profile until time $t$.  For lumber degradation, we assume two basic properties for $\eta_t$:
\begin{itemize}
\item[(i)] If $\tau(t) \le \tau^*$ for $\delta_1 \le t \le \delta_2$, then $\dot{\eta}_t = 0 \mbox{~~} \forall \mbox{~~} t \in (\delta_1, \delta_2)$, where $\tau^*$ is a threshold stress level below which the population does not undergo degradation.
\item[(ii)] If $\tau(t)$ is held at a constant level larger than $\tau^*$ for $\delta_1 \le t \le \delta_2$, then $\dot{\eta}_t$ is decreasing over the interval $\delta_1 \le t \le \delta_2$.
\end{itemize}
The first property implies that degradation does not progress during periods when the stress is too low to cause damage.  The threshold $\tau^*$ is a population analogue of the damage threshold commonly seen in ADMs (see Introduction).  The second property captures the DOL effect:  if the load is held constant at a stress level high enough to cause failures in the population, degradation continues as that constant load is maintained but the rate at which it occurs is expected to slow over time.  These properties will guide the specific choice of $\eta_t$.

\subsection{A model for the shape parameter}

We now develop a specific functional form for the shape parameter $\eta_t$ along with parameters to be estimated from data in the illustrative example.  The ``power law'' and its variants have been commonly used to model degradation and serves as a useful starting point for developing specific model implementations.

Suppose $\tau$ is a given load level held constant over time.  Then we can conceive a simple form for $\eta_t$ to characterize the degradation in a population of pieces subject to that load from time $0$ to $t$ as
\begin{eqnarray}\label{eq:basicDOL}
\eta_t = g(t) \times ( u \tau - v)_+,
\end{eqnarray}
where $g(\cdot)$ is an increasing function that captures the DOL effect, $u$ and $v$ are positive constants, and $x_+ = \max(x,0)$. Here the term $(u \tau  - v)_+$ is constant over time, depending only on the size of the load.  It is zero when that stress level is sufficiently low in accord with property (i), that is, when $u \tau_i  - v < 0$ which corresponds to the stress threshold $\tau^* = v/u$ below which no degradation occurs.  The function $g(t)$ governs the rate of degradation in the population over time under that fixed load.  The simple form $g(t) = t^a$ with  $a>0$ would reproduce the well-known power law.  Various modifications can be made to increase its flexibility to model the degradation behaviour, and we will perform our subsequent analysis using the form $g(t) = t^a + bt^c$ where $a,b,c$ are all positive parameters with $a < c$, which has the feature of mixing two different power law growth rates.  In particular by setting the constraint $a<c$ we expect that $t^a$ will capture the shorter-term effect well, while the role of $bt^c$ becomes more important over longer time durations.

In practice, the load may vary over time.  Let $0 = \tau_0 < \tau_1 < \tau_2 < \cdots < \tau_m$ denote a sequence of load levels spanning the range of loads under which the population may be subjected. Then for each load level $\tau_i$, $i = 1, \ldots, m$, we can consider the amount of incremental degradation due to load $\tau_i$ beyond that which was sustained from load $\tau_{i-1}$.  Then a natural analogue to Equation (\ref{eq:basicDOL}) for this load increment, for time 0 to $t$, is
\begin{eqnarray*}
g(\tilde{t_i}) \left[(u\tau_i -v)_+ - (u \tau_{i-1}-v)_+ \right],
\end{eqnarray*}
where $\tilde{t_i} = \int_0^t I(\tau(t') \ge \tau_i) \, dt'$ is the total time duration for which the load exceeded $\tau_i$.  Thus the constant term $\left[(u\tau_i -v)_+ - (u \tau_{i-1}-v)_+ \right]$ captures the incremental `jump' in $\eta_t$ that occurs due to load level $\tau_i$ being reached.  Similarly $g(\tilde{t_i})$, as a function of the total length of time for which the load level $\tau_i$ is sustained, now models its corresponding DOL effect.

We can then combine the contributions of all the load levels to construct $\eta_t$ for any arbitrary given load profile.  Using our chosen form for $g(t)$, we thus obtain
\begin{eqnarray}\label{eq:eta_t}
\eta_t = \sum_{i=1}^m (\tilde{t_i}^a + b \tilde{t_i}^c) \left[(u\tau_i -v)_+ - (u \tau_{i-1}-v)_+ \right],
\end{eqnarray}
which reduces to Equation (\ref{eq:basicDOL}) in the special case that the load is held constant at $\tau$ from time $0$ to $t$.  It can be seen that $\eta_t$ is differentiable, since if  $\tau_{j} \le \tau(t) < \tau_{j+1}$, we have
\begin{eqnarray}
\dot{\eta}_t = \sum_{i=1}^j (a \tilde{t_i}^{a-1} + bc \tilde{t_i}^{c-1}) \left[(u\tau_i -v)_+ - (u \tau_{i-1}-v)_+ \right].
\end{eqnarray}
Thus when the exponents $a$ and $c$ are each less than 1, $\dot{\eta}_t$ is decreasing over any period with a fixed load level, in accord with property (ii).

A specific sequence of load levels $\tau_1, \ldots, \tau_m$ needs to be chosen for computation.  These serve as the incremental thresholds over which additional degradation contributions are added into the model.  For example, if $\tau_j = 3000$psi and $\tau_{j+1} = 3020$psi, then any loads in the interval $[3000,3020)$ would contribute the same amount to $\eta_t$ in this model as a load of exactly 3000psi.  Naturally, the range of loads may be discretized as finely as desired to faithfully reproduce the stress history, at the cost of additional computation time.  In our demonstration we use an equally-spaced sequence for $\tau_1, \ldots, \tau_m$  with intervals of 20psi.  An artifact of the discrete load levels in the model is that if the load profile $\tau(t)$ has periods of continuous increase, the resulting $\eta_t$ becomes jagged as the load passes the different thresholds rather than smoothly increasing with the load.  In this case a line segment can be used to smooth $\eta_t$ between the time points when successive load thresholds are reached, to serve as an acceptable approximation.

\section{A Bayesian analysis of degradation}\label{sect:BayesianAnalysis}
This section presents a Bayesian analysis of data
from an accelerated testing experiment designed to explore the duration of load effect.

\subsection{The data}\label{sect:data}
The real data we subsequently analyze come from the DOL experiment reported in \citet{foschi1982load}.  It consists of a total of 637 pieces of visually graded 2x6 Western Hemlock, divided for testing under three different load profiles (all time units in hours unless otherwise indicated):
\begin{enumerate}
\item 198 pieces were assigned the load profile
\begin{equation*}\label{eq:loadprof}
\tau(t) = \begin{cases}
388440t, \mbox{~~~~for } t \le 3000/388440 \\
4500, \mbox{~~~~for }  3000/388440 < t \le 4\mbox{~years}, \end{cases}
\end{equation*}
i.e., the load was increased linearly until reaching 3000psi, and held at that constant level for 4 years.  Hence pieces that do not fail by the end of the 4-year period when the test is truncated have their failure time censored.
\item 300 pieces were assigned the load profile
\begin{equation*}
\tau(t) = \begin{cases}
388440t, \mbox{~~~~for } t \le 4500/388440 \\
4500, \mbox{~~~~for }  4500/388440 < t \le 1\mbox{~year}, \end{cases}
\end{equation*}
which is similar to the above, now with a constant load level of 4500psi for 1 year.  Pieces that do not fail by the end of the 1-year period when the test is truncated have their failure time censored.
\item 139 pieces were assigned the load profile $\tau(t) = 388440t$ until failure.
\end{enumerate}
In the DOL literature, profiles 1 and 2 are known as `constant load' tests, while profile 3 is known as a `ramp load' test.  These are so-called `accelerated' testing schemes that were originally designed to help elucidate the long-term DOL effect using tests of relatively shorter duration \citep{barrett1978durationII}.

Each piece that failed during the test had its failure time recorded.  Pieces that did not fail during the test duration had their censoring times recorded (i.e., 4 years for group 1 and 1 year for group 2).  No covariates for individual specimens were recorded in the data.

\subsection{Fitting the  degradation model}\label{sec:modelfitting}
We now perform an illustrative analysis of these accelerated testing data based on the model developed, using the techniques of Bayesian inference.  Let $\theta$ denote the vector of parameters to be inferred, which consists of the five parameters associated with the model for $\eta_t$ along with the gamma process scale parameter $\xi$, namely $\theta = (a,b,c,u,v,\xi)$.  Let $\pi(\theta)$ denote the joint prior distribution on $\theta$.  Then using the likelihood in Equation (\ref{eq:likT}), the posterior distribution of $\theta$ based on an independent sample of test specimens with recorded failure times $t_1, t_2, \ldots, t_n$ is given by
\begin{eqnarray}\label{eq:posterior}
\pi ( \theta | t_1, t_2, \ldots, t_n) \propto \pi(\theta) \prod_{i=1}^n f_T( t_i | \xi, \eta_{t_i} ),
\end{eqnarray}
where $\eta_{t_i}$ denotes evaluating Equation (\ref{eq:eta_t}) for $\eta_t$ at time $t_i$ according to the load profile $\tau(t)$ associated with specimen $i$.  For some specimens the actual failure times are not observed, as the test has ended after a specified duration without the specimen failing.  Then the likelihood contribution $f_T( t_i | \xi, \eta_{t_i} )$ for those specimens is replaced by the corresponding survivor function, namely $P( T_i > t_c | \xi, \eta_{t_i} )$ computed by Equation (\ref{eq:surv}) where $t_c$ is the truncation time.

Equation (\ref{eq:posterior}) thus can accommodate all the test data to be analyzed under the different loading profiles employed in the experiment.  Importantly, we emphasize it is assumed that the same set of parameters can model the degradation of the population under \emph{any} loading scenario.  That assumption, which implies that the parameters of a fitted model can then be used with any load profile $\tau(t)$ of interest, has been fundamental to much of the previous work with ADMs that involve the probabilistic assessment of long-term lumber reliability.  An example of such follows in Section \ref{sect:reliabilityanalysis}.

To proceed with the analysis, we use vague independent Normal($\mu=0$, $\sigma=1000$) priors for each of the parameters in $\theta$, along with the restriction $a<c$. As the form of the posterior is intractable for direct sampling, we employ Markov Chain Monte Carlo (MCMC) techniques to obtain sample draws from it.  To obtain reasonable starting values for the MCMC, we first used Nelder--Mead iterations to optimize the posterior.  Then, to improve convergence and the efficiency of posterior exploration via MCMC, we used parallel tempering \citep{swendsen1986replica} distributed over 120 compute cores, with each core running a MCMC chain using simple Metropolis-Hastings iterations and temperatures geometrically spaced from 1 to 20.  Swaps between chains were performed every five iterations.  The first 5,000 iterations were discarded as burn-in, and the following 15,000 iterations from the chain representing the target posterior distribution constitute our final samples.

Summaries of the posterior samples of the parameters are shown in Table \ref{tab:fitted}.  A few observations can be noted.  First, there is a clear distinction between the powers $a$ and $c$, with posterior means of 0.019 and 0.40 respectively, indicating that a single power law does not adequately explain the observed degradation over time.  Second, there is only weak evidence for a stress threshold $\tau^*$ below which no population degradation occurs; the MCMC samples yield a posterior mean for the threshold level $v/u$ of 413psi and a highly uncertain 95\% posterior interval $(43,642)$ so that a very low threshold is plausible.  Third, the highest uncertainty is in the parameter $b$, whose central 95\% posterior probability interval spans two orders of magnitude:  $(0.00729, 0.03732)$.  This indicates that the true degradation behaviour over longer time durations (i.e., a year or more) is highly uncertain from these data alone, with the two constant load tests having been truncated at 1 and 4 years.

\begin{table}[htbp]
  \centering
  \caption{Summary of the posterior distributions of the parameters in the fitted gamma process model.}
    \begin{tabular}{r|rrr|r}
          Parameter & \multicolumn{3}{c|}{Posterior quantiles} & Posterior mean  \\
           & 50\%  & 2.50\% & 97.50\%   \\
          \hline
    $a$ & 0.019  & 0.012  & 0.027  & 0.019 \\
    $b$ & 0.00729  & 0.00071  & 0.03732 & 0.01026 \\
    $c$ & 0.39 & 0.25 & 0.60 & 0.40 \\
    $u$ & 0.00088  & 0.00071  & 0.00108 & 0.00088 \\
    $v$ & 0.388 & 0.041 & 0.584 & 0.359 \\
    $\xi$ & 0.21  & 0.16  & 0.26 & 0.21
    \end{tabular}%
  \label{tab:fitted}%
\end{table}%

\begin{figure}[!htbp]
\centering
\includegraphics[scale=0.62,angle=-90]{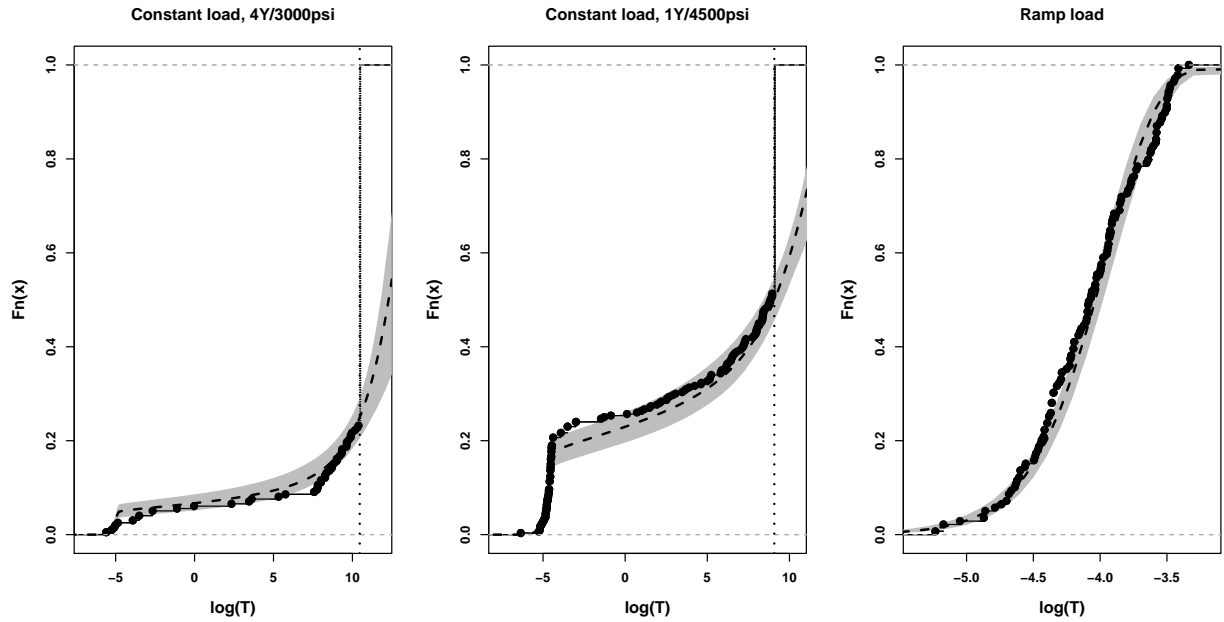}
\caption{Cumulative distribution functions (CDFs) associated with the fitted gamma process model, on the three calibration datasets.  In each plot, the black points show the empirical CDF of the dataset.  The dashed curve is the CDF associated with the set of parameters with the highest posterior density among the MCMC samples, while the grey area represents the 95\% posterior probability interval of the CDF based on the MCMC samples. The vertical dotted lines indicate the censoring times for the two constant load scenarios.}
\label{fig:cdfs}
\end{figure}

The proposed model fits the data for the three test scenarios well, as can be seen in the plots in Figure \ref{fig:cdfs}.  The cumulative distribution functions (CDFs) computed from the sampled parameter vectors largely capture the empirical distributions.  The 95\% posterior bands, shown in grey, are also tight for the time ranges over which failures are observed.  Beyond the test truncation times, namely 4 years for the 3000psi constant load group and 1 year for the 4500psi constant load group, the uncertainty increases  substantively as seen in the width of the posterior intervals.  Hence projections of degradation over the long term, say 30 or 50 years, based on these data alone would likewise have very high variability.

\section{Reliability analysis: an illustrative example}\label{sect:reliabilityanalysis}

\begin{figure}[!htbp]
\centering
\includegraphics[scale=0.62,angle=-90]{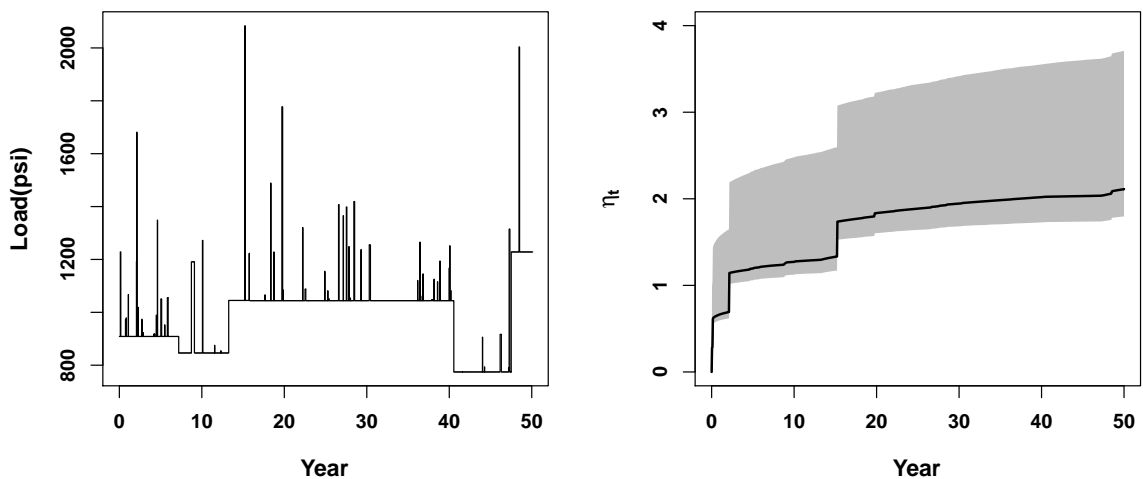}
\caption{Reliability assessment example.  The left panel shows an example of a simulated residential 50-year load profile, adapted from \citet{foschi1989reliability}. The right panel shows the corresponding $\eta_t$ of the fitted gamma process model under this load profile.  The black curve shown is computed on the set of parameters with the highest posterior density among the MCMC samples, while the grey area represents the 95\% posterior probability interval based on the MCMC samples.}
\label{fig:load}
\end{figure}

We now turn to applying the fitted model to an example of a predictive scenario, such as those analyzed in reliability assessments.  \citet{foschi1989reliability} use stochastic processes to characterize load profiles on individual lumber members over the lifetime of a wood structure, and an adapted example of a heavier than typical 50-year load profile for a residential dwelling unit is shown in the left panel of Figure \ref{fig:load}.  This profile is a piecewise constant function obtained by summing different component loads.  Intuitively, the total load at any given time includes the constant dead weight of the structure, along with load from occupancy which varies by resident.  In addition, the `spikes' correspond to various short-term loads that are expected to occur periodically in homes.

Using the parameters from the fitted model, we may compute $\eta_t$ corresponding to this load profile using Equation (\ref{eq:eta_t}).  The solid black curve shows $\eta_t$ computed for this 50-year period using the sampled parameter vector with the highest posterior density.  It can be seen that $\eta_t$ increases rapidly the first time the load exceeds a new threshold, for example, at time $\sim$2 years (load $\sim$1675psi) and $\sim$15 years (load $\sim$2050psi).  Subsequent loadings translate to more modest degradation increases over time, as expected from the DOL effect; for example, the second time the load exceeds 2000psi at time $\sim$48 years its effect on $\eta_t$ is much more diminished.  As before, the grey area represents 95\% posterior bands based on the MCMC samples.

Ultimately the probability of failure by the end of the 50-year period is of primary interest.  This is determined by the value of $\eta_t$ at 50 years, along with the scale parameter $\xi$ of the gamma process according to Equation (\ref{eq:surv}).  We obtain the posterior mean for the probability of failure of 0.090, and a central 95\% posterior interval of $(0.055,0.150)$.

The reliability calculations based on the gamma process model are fast and simple, compared to the ADM approach which requires numerically solving an ODE for a large number of simulated pieces to estimate the probability of failure.  To compare results, the approach of \citet{foschi1989reliability} based on the Canadian ADM and their parameter estimates from these same data, yield a 50-year failure probability of 0.015 for this load profile.  Thus there is a large discrepancy between the long-term predictions from the different approaches, even though both approaches are able to fit the empirical data quite well.  However Foschi's approach does not provide for the construction of confidence intervals to assess uncertainty.  We comment on this issue further in the discussion section.

\section{Predicting the residual life of lumber in service}\label{sect:residuallife}

As a further application of the fitted Bayesian model, we may use the MCMC samples to compute the posterior probability distributions of the residual life for pieces that have not failed up to a given time $t'$.  This requires a knowledge of $\eta_{t'}$, which in our model is computed from the load profile $\tau(t); 0 \le t \le t'$ and the fitted parameters, as well as a characterization of the expected future loads $\tau(t); t>t'$.  Letting  $T$ denote the random variable for the failure time, then of interest is the distribution of $T_r :=  [ T | T > t' ] - t'$ which represents the remaining lifetime.  It has survivor function
\begin{eqnarray}\nonumber
P [T_r > t_r \mid \xi,\eta_t, \eta_{t'}] &=&  \frac{ P [T > t' + t_r \mid \xi,\eta_t] }{ P[T > t' \mid \xi,\eta_{t'}] },
\end{eqnarray}
which may be computed using Equation (\ref{eq:surv}).

\begin{figure}[!htbp]
\centering
\includegraphics[scale=0.62,angle=-90]{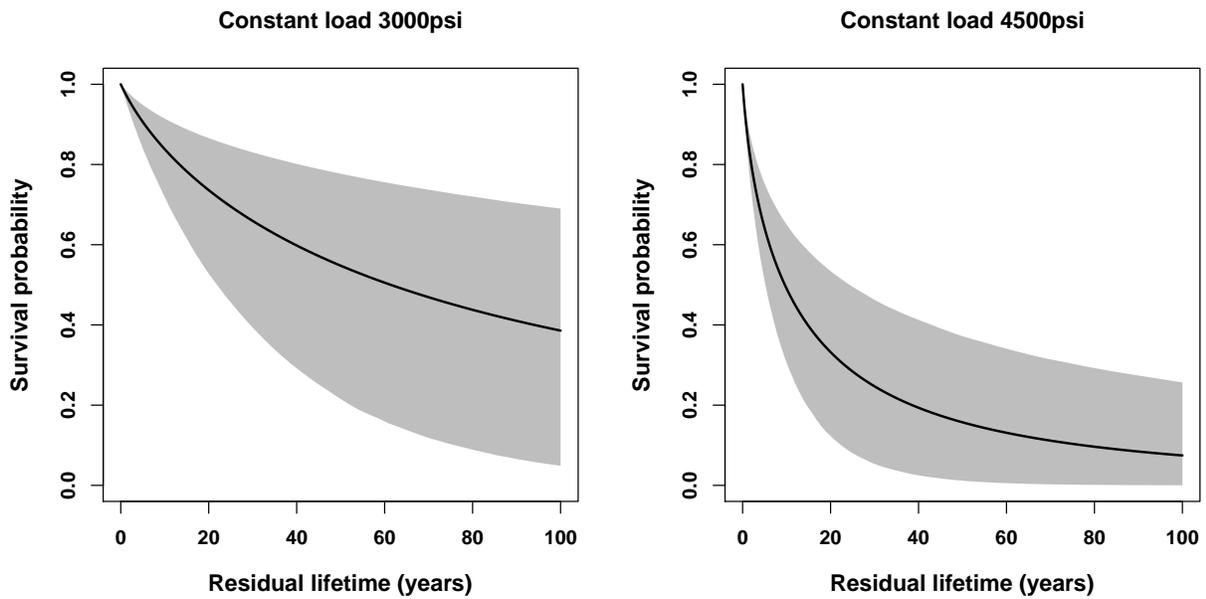}
\caption{Residual lifetime example.  Survivor functions of the remaining lifetime for specimens under a continued 3000psi constant load after surviving the 4-year test period (left panel), and for specimens under a continued 4500psi constant load after surviving the 1-year test period (right panel).  The black curve shown is the posterior mean, while the grey area represents the 95\% posterior interval based on the MCMC samples.}
\label{fig:resid}
\end{figure}

To illustrate, we use the two constant-load scenarios in the experimental data, where specimens were held at load levels of 3000psi and 4500psi for 4 years and 1 year respectively.  Consider the distribution of remaining lifetime of the surviving specimens, if these constant load levels were maintained indefinitely.  These survivor functions are shown, for up to 100 more years, in Figure \ref{fig:resid}, with posterior uncertainty shown by the grey bands.  These distributions have very long right tails, corresponding to the strongest members of the population which can carry these load levels almost indefinitely.  As such, the mean residual lifetime is not very meaningful.  Instead quantities such as the time until 50\% of the survivors fail, namely the median of these distributions, may be of interest.  Using the MCMC samples, we calculate the 95\% posterior intervals of these medians to be $(21.9, 333.5)$ years under 3000psi and  $(5.2,24.3)$ years under 4500psi.  It can be seen that there is much higher uncertainty associated with these distributions at the lower load level.

\section{Discussion and conclusions}\label{sect:discussion}

In the analysis of the experimental data we found that the effect of degradation from a constant load due to time, as modeled in the shape parameter, was not a power law $t^a$.  This is evident by examining the plots in Figure \ref{fig:cdfs}.  With a simple power law, the CDF would be approximately linear as a function of log-time during the constant load period. Instead, the empirical CDF increases quite nonlinearly with time on the log-scale.  This led us to posit adding a second power term to the model, yielding $t^a + bt^c$ with $a < c$.  This form provides a good fit to the data, however with wide posterior intervals for the parameters $b$ and $c$.  That in turn translates to the high uncertainty that we find associated with using tests of 1 and 4 year durations to predict reliability and residual lifetime over much longer periods, such as 50 years.  Larger tests, or over longer periods, would be necessary to reduce this variability.

In the work by \citet{foschi1989reliability}, a crucial parameter in the Canadian ADM used for reliability analysis is the `stress threshold' $\sigma_0$. In that model it is hypothesized that an individual piece of lumber does not accumulate damage when the load is below $\sigma_0 \tau_s$, where $\tau_s$ is the strength of that piece as measured in a short-term ramp load test.  That work reported an estimate for the population mean of $\sigma_0$ to be 0.533; based on that estimate along with a population mean short-term strength of $\sim$6900psi, most pieces do not eventually fail under the load levels seen in the residential example of Figure \ref{fig:load}.  However, in subsequent re-analysis of that model based on the same data, the population mean for $\sigma_0$ was found to be highly uncertain and a strong Bayesian prior was needed to stabilize its estimate \citep{yang2017adm}; in fact, a mean of $\sigma_0 \approx 0$ can still fit the empirical data well by adjusting the other ADM parameters.  Hence, when such parameter uncertainty is accounted for, the ADM approach likewise would yield wide prediction intervals.  It may well be that the estimate of 0.533 reflects some other, not explicitly reported prior knowledge about the behaviour of lumber, e.g.~how many wood structures have survived the test of 50 or 100 years.  However in the current application no information concerning that issue was available. In the context of the gamma process approach, such information could easily be incorporated into the priors for Bayesian analysis, to set more realistic constraints on the rate of degradation over longer periods.

We would further note that $\sigma_0$ as a piece-level parameter in the ADM does not have a direct relationship with our estimated damage threshold of 413psi for the population.  In the ADM, the population mean of $\sigma_0 \tau_s$ is the load below which the \emph{average} piece in the population is undamaged; however, the realization of $\sigma_0 \tau_s$ cannot be assessed for any individual piece since it is unobservable.  In contrast the 413psi population threshold in our model represents the stress level below which \emph{all} members of the population are undamaged.  Nonetheless as discussed above, both approaches show little evidence of a high damage threshold by analyzing the Hemlock data alone, when uncertainty is considered.  Specialized proof-loading tests \citep[e.g.,][]{woeste2007proof} may instead be more useful if estimating the damage threshold is of primary interest.

Another point of comparison between the ADM and our proposed approach lies in the number of parameters to be estimated.  Fitting the Canadian ADM in particular requires estimating 10 population parameters (the five log-normal means and variances from which the random effects in Equation (\ref{eq:canadianmodel}) are drawn for specific pieces of lumber), some of which do not have a clear physical interpretation.  As found in \cite{yang2017adm}, a number of different sets of these population parameters could lead to essentially the same likelihood, suggesting that while the Canadian ADM can fit the empirical data well, it may be over-parametrized leading to worse prediction performance due to the inflated uncertainty about the individual parameters.  Our model fits the empirical data well with four fewer parameters (six), and it is simpler to see that the resulting uncertainty in prediction stems primarily from the uncertainty in the estimation of degradation rate over longer periods based on accelerated testing data.

It can be said that the results of applying the accumulated damage modeling approach along with its predecessor, the empirical model of \citet{wood1960relation}, have laid a foundation for incorporating long term stress effects into the calculation of design values that have stood the test of time. So why a critical review of these models at this time?  The answer lies in the need for application of the methods to a new generation of forest products such as strand based wood composites \citep{wang2012doltheory,wang2012dolexperiment} that are also susceptible to DOL effects.  Given that the new applications do not automatically inherit the record of success of the ADM, prudence suggests a re-evaluation of the approach given its limitations as described in the Introduction, one that takes full advantage of the new computational and statistical methods now available.  Since engineered wood composites have much lower short-term strength variability compared to lumber, the size of the DOL effect (and its estimation) for these materials would have a more significant role in determining appropriate safety factors.

The above considerations led the authors to explore the alternative to the ADM presented in this paper and it was found to overcome many of the difficulties described above with the ADM approach.  The model based on the gamma process is simpler to interpret with fewer parameters, separates external (population) and internal (individual piece) sources of variability, and lends itself well to standard statistical assessments of uncertainty.  The degradation approach also led to a number of new discoveries as previously summarized in the Introduction.  In particular, a key finding from our analysis is that the accelerated testing data yields poor predictors of the long term future of a piece of lumber in service.  Our analysis shows very wide credibility bands for the median time to failure, particularly when the sustained load level for the test is low.  This finding suggests much larger accelerated tests are needed to ensure the reliability of predictions.

\subsection*{Acknowledgements}
The work reported in this manuscript was partially supported by FPInnovations and a CRD grant from the Natural Sciences and Engineering Research Council of Canada.  The data analysed in this paper were provided by FPInnovations. We are greatly indebted to Conroy Lum and Erol Karacabeyli from FPInnovations for their extensive advice during the conduct of the research reported herein.

\bibliographystyle{apalike}

\begin{thebibliography}{}

\bibitem[Barrett and Foschi, 1978a]{barrett1978durationI}
Barrett, J. and Foschi, R. (1978a).
\newblock Duration of load and probability of failure in wood. part i.
  modelling creep rupture.
\newblock {\em Canadian Journal of Civil Engineering}, 5(4):505--514.

\bibitem[Barrett and Foschi, 1978b]{barrett1978durationII}
Barrett, J. and Foschi, R. (1978b).
\newblock Duration of load and probability of failure in wood. part ii.
  constant, ramp, and cyclic loadings.
\newblock {\em Canadian Journal of Civil Engineering}, 5(4):515--532.

\bibitem[Ellingwood and Rosowsky, 1991]{ellingwood1991duration}
Ellingwood, B. and Rosowsky, D. (1991).
\newblock Duration of load effects in lrfd for wood construction.
\newblock {\em Journal of Structural Engineering}, 117(2):584--599.

\bibitem[Foschi, 1984]{foschi1984reliability}
Foschi, R.~O. (1984).
\newblock Reliability of wood structural systems.
\newblock {\em Journal of Structural Engineering}, 110(12):2995--3013.

\bibitem[Foschi and Barrett, 1982]{foschi1982load}
Foschi, R.~O. and Barrett, J.~D. (1982).
\newblock Load-duration effects in western hemlock lumber.
\newblock {\em Journal of the Structural Division}, 108:1494--1510.

\bibitem[Foschi et~al., 1989]{foschi1989reliability}
Foschi, R.~O., Folz, B., and Yao, F. (1989).
\newblock {\em Reliability-based design of wood structures}.
\newblock Number~34. Dept. of Civil Engineering, University of British
  Columbia.

\bibitem[Gerhards, 1979]{gerhards1979time}
Gerhards, C.~C. (1979).
\newblock Time-related effects on wood strength: A linear cumulative damage
  theory.
\newblock {\em Wood science}, 11:139--144.

\bibitem[Haupt, 1867]{haupt1867general}
Haupt, H. (1867).
\newblock {\em General theory of bridge construction}.
\newblock Appleton, New York.

\bibitem[Hoffmeyer and S{\o}rensen, 2007]{hoffmeyer2007duration}
Hoffmeyer, P. and S{\o}rensen, J.~D. (2007).
\newblock Duration of load revisited.
\newblock {\em Wood Science and Technology}, 41(8):687--711.

\bibitem[Karacabeyli and Soltis, 1991]{karacabeyli1991state}
Karacabeyli, E. and Soltis, L.~A. (1991).
\newblock State of the art report on duration of load research for lumber in
  north america.
\newblock In {\em Proceedings of the 1991 International Timber Engineering
  Conference. London, United Kingdom}.

\bibitem[Lawless and Crowder, 2004]{lawless2004covariates}
Lawless, J. and Crowder, M. (2004).
\newblock Covariates and random effects in a gamma process model with
  application to degradation and failure.
\newblock {\em Lifetime Data Analysis}, 10(3):213--227.

\bibitem[Paroissin and Salami, 2014]{paroissin2014failure}
Paroissin, C. and Salami, A. (2014).
\newblock Failure time of non homogeneous gamma process.
\newblock {\em Communications in Statistics-Theory and Methods},
  43(15):3148--3161.

\bibitem[Rosowsky and Bulleit, 2002]{rosowsky2002another}
Rosowsky, D.~V. and Bulleit, W.~M. (2002).
\newblock Another look at load duration effects in wood.
\newblock {\em Journal of Structural Engineering}, 128(6):824--828.

\bibitem[Swendsen and Wang, 1986]{swendsen1986replica}
Swendsen, R.~H. and Wang, J.-S. (1986).
\newblock Replica monte carlo simulation of spin-glasses.
\newblock {\em Physical Review Letters}, 57(21):2607.

\bibitem[Wang et~al., 2012a]{wang2012doltheory}
Wang, J.~B., Foschi, R.~O., and Lam, F. (2012a).
\newblock Duration-of-load and creep effects in strand-based wood composite: a
  creep-rupture model.
\newblock {\em Wood science and technology}, 46(1-3):375--391.

\bibitem[Wang et~al., 2012b]{wang2012dolexperiment}
Wang, J.~B., Lam, F., and Foschi, R.~O. (2012b).
\newblock Duration-of-load and creep effects in strand-based wood composite:
  experimental research.
\newblock {\em Wood science and technology}, 46(1-3):361--373.

\bibitem[Woeste et~al., 2007]{woeste2007proof}
Woeste, F., Green, D., Tarbell, K., and Marin, L. (2007).
\newblock Proof loading to assure lumber strength.
\newblock {\em Wood and fiber science}, 19(3):283--297.

\bibitem[Wong and Zidek, 2016]{wong2016dimension}
Wong, S.~W. and Zidek, J.~V. (2016).
\newblock Dimensional and statistical foundations for accumulated damage
  models.
\newblock {\em arXiv preprint arXiv:1708.03018}.

\bibitem[Wood et~al., 1960]{wood1960relation}
Wood, L.~W. et~al. (1960).
\newblock {\em Relation of strength of wood to duration of load}.
\newblock Madison, Wis.: US Dept. of Agriculture, Forest Service, Forest
  Products Laboratory.

\bibitem[Yang et~al., 2017]{yang2017adm}
Yang, C.-H., Zidek, J.~V., and Wong, S.~W. (2017).
\newblock Bayesian analysis of accumulated damage models in lumber reliability.
\newblock {\em arXiv preprint arXiv:1706.04643}.

\bibitem[Zhai, 2011]{zhai2011dynamic}
Zhai, Y. (2011).
\newblock Dynamic duration of load models.
\newblock Master's thesis, University of British Columbia, Department of
  Statistics.

\bibitem[Zhai et~al., 2012]{zhai2012review}
Zhai, Y., Pirvu, C., Heckman, N., Lum, C., Wu, L., and Zidek, J.~V. (2012).
\newblock A review of dynamic duration of load models for lumber strength.
\newblock Technical report, TR 270, Department of Statistics, University of
  British Columbia.

\end{thebibliography}

\end{document}